\documentclass{article}
\usepackage[utf8]{inputenc}

\usepackage[T1]{fontenc}
\usepackage{times}%

\usepackage{amsmath}
\usepackage{dcolumn}
\usepackage{graphics}

\usepackage{graphicx}
\usepackage{subfig}
\usepackage{color}
\usepackage{booktabs}
\usepackage{multirow}
\usepackage{enumitem}  
\usepackage{hyperref}

\usepackage{stfloats}

\usepackage{authblk}
\usepackage{blindtext}

\title{Predicting consumers engagement on Facebook based on \textit{what} and \textit{how} companies write
}

\author[1]{Érika S. Rosas-Quezada}
\author[1]{Gabriela Ramírez-de-la-Rosa} 
\author[1,2]{Esaú Villatoro-Tello\thanks{Corresponding author.}}

\affil[1]{Language and Reasoning Research Group \protect\\
Information Technologies Department\protect\\
Universidad Aut\'onoma Metropolitana - Cuajimalpa, M\'exico \protect\\
{\textit{eye.erika.rosas@gmail.com, \protect\\ \{gramirez, evillatoro\}@correo.cua.uam.mx}}
}

\affil[2]{Idiap Research Institute \protect\\ 
Rue Marconi 19, 1920 Martigny, Switzerland \protect\\
\textit{esau.villatoro@idiap.ch}
}

\date{September 2019}

\begin{document}

\maketitle

\begin{abstract}
Engaged costumers are a very import part of current social media marketing. Public figures and brands have to be very careful about what to post online. That is why the need for accurate strategies for anticipating the impact of a post written for an online audience is critical to any public brand. Therefore, in this paper, we propose a method to predict the impact of a given post by accounting for the content, style, and behavioral attributes as well as metadata information. For validating our method we collected Facebook posts from 10 public pages, we performed experiments with almost 14000 posts and found that the content and the behavioral attributes from posts provide relevant information to our prediction model. 

\end{abstract}

\section{Introduction}
\label{sect:Introduction}
Nowadays, people worldwide are largely engaged and attached to different types of Internet technologies and Social media platforms. All these technologies combined have provided new ways for exchanging feedback on products and services. As stated in \cite{moro2018brand}, this type of circumstances has boosted customer empowerment. Accordingly, customers have the potential of becoming influential with their opinions, recommendations or complaints. 

This situation requires the constant incorporation of novel strategies for effectively managing brand's aims and marketing plans, especially aspects related to customers' involvement, relationship, and communication \cite{alalwan2017social}. Thus, measuring the impact of produced advertising is an important issue that needs to be included by brands as part of their social media management strategies \cite{moro2016predicting}. According to previous research, the impact of a published post is measured through several available metrics, mainly related to the consumer's visualizations, reactions, comments, and interactions. Hence, increasing the impact of the published posts will lead to stronger relationships among brand and consumers, allowing customers to create valuable content through social media \cite{schultz2017proposing}.  

Recently, the community of electronic commerce and business research has started to pay attention to how effectively exploit the mechanisms to interact with their customers. Researches have focused on studying phenomena such as the role of social media on advertising, the electronic word of mouth, customer's relationships management, brand's performance, among others \cite{alalwan2017social, amado2018research, schreiner2019impact}. Although many works have proposed techniques for finding the relationships between online posts on social media and the impact of such publications measured by users interactions, the vast majority of these research do it as a posteriori analysis \cite{alalwan2017social,amado2018research,bonson2015citizens, gao2016branding, moro2018brand,schultz2017proposing}. This means they focus on finding those characteristics that allowed a post to be appealing for their customers, obtaining valuable insights that enable designing powerful marketing strategies. However, in spite of all the knowledge that these methodologies can provide to specific firms, it is not enough for predicting the impact a post will have prior to its publication. Therefore, a system able to anticipate the impact of individual posts can provide an enormous advantage when deciding to communicate something to the costumers through social media platforms.

In this paper, we propose a novel framework for predicting the impact of publishing posts on a social media network, namely Facebook. Contrastingly to traditional approaches in the field, our method incorporates features that are able to capture content, style, and behavioral features when representing posts. The proposed approach is based on a supervised machine learning strategy, which allows anticipating post's impact, i.e., either high- or low-impact.  For validating the proposed method, we took on the task of collecting a dataset from ten renowned brands on Facebook Mexico. Our performed experiments, over more than 13,000 posts, for six different classification problems, indicate that the combination of the proposed features with some metadata-based attributes, allows an automatic system to obtain acceptable performance results.

We foresee this work will represent an important contribution to the development of novel methodologies in the field of electronic commerce and business research, as well as motivate further research from the intelligent systems and text mining research communities.

The main contributions of this paper are as follows:

\begin{enumerate}
    \item We collected and labeled more than 14,000 posts from ten renowned brands on Facebook Mexico. This dataset represents a valuable resource for future research work on the field of electronic commerce and business, as well as for the intelligent systems community.
    
    \item We provide evidence on the importance of content-based, stylistic, and behavioral features in combination with metadata-based attributes for solving the task of impact prediction on Facebook posts.

    \item We proposed a novel framework, based on a supervised machine learning approach, for solving the problem of anticipating the impact of publishing a post on Facebook. 
\end{enumerate}

The rest of the paper is organized as follows. The next section provides a review of related work on the problem of social media and customer relationships management.  Section \ref{sect:corpus} describes the followed methodology for collecting the employed dataset, how it was labeled, and provides some statistics regarding its composition. Section \ref{sec:proposedframework} explains the proposed framework based on an supervised approach for predicting the impact of publishing posts on Facebook. Section \ref{sec:experiments} depicts the experimental setup, and the obtained results for all the performed experiments. Finally, in Section \ref{sec:conclusions} we draw some conclusions and future work directions.

\section{Related work}
\label{sect:relatedWork}

Consumer engagement is measured by the number of performed activities by users within the social media platform. Normally, these activities vary from platform to platform, however, on Facebook, a typical set of metrics that help to evaluate the level of engagement are: generated reactions (positive, negative, and neutral reactions), number of comments, and the number of times a post is shared \cite{schultz2017proposing}. Thus, posts having elevated or low numbers under these metrics, are considered examples of high or low impact posts respectively; meaning a healthy/unhealthy customer engagement relationships.  

Accordingly, literature establishes that the more capable are the organizations building and sustaining emotional and social ties between their customers and their brands, i.e., a healthy level of customer engagement; the more the benefits that can be obtained. Therefore, many research groups have tackled the problem of how to contribute to both customers experience and customer relationships using social media platforms \cite{alalwan2017social, amado2018research}. 

On the one hand, the vast majority of the previous work has faced the problem as a knowledge extraction technique for designing powerful marketing strategies. In other words, this type of research proposes analyzing the relationships between several variables and the level of engagement of customers. Thus, it is possible to find what are the main characteristics that provoke customers manifestations (reactions, comments, and sharing). However, a major drawback of these approaches is that they do not consider using this knowledge as part of an automatic method for anticipating the impact of a post. Recent examples of this type of methodologies can be found in \cite{alalwan2017social,amado2018research,bonson2015citizens, gao2016branding, moro2018brand,schultz2017proposing}. 

On the other hand, a few research works have proposed and evaluated distinct methodologies for implementing predictive systems \cite{moro2016predicting, sabate2014factors, silva2018unveiling, yano2010s}. In \cite{moro2016predicting} authors proposed using seven features for representing the information contained in a post, namely: category of the post (action, product, or inspirational), the total likes of the brand's page, the type of content (photo, video, or link), time of the publication, month, weekday and hour of the post, and a feature that indicates if the post was paid for advertising. These features were employed for predicting 12 distinct Facebook metrics. 
For their experiments, authors employed a SVM regresor, and evaluated their method in 790 posts from a cosmetic company's page. A similar approach is described in \cite{silva2018unveiling} but for estimating the success of eBay smartphone sellers. For representing the data authors proposed near 20 metadata-based features extracted from the eBay platform, such as reachability and engagement (followers), customer feedback (number of positive and negative reviews) and seller information (name, country, etc.). In the work of \cite{sabate2014factors}, 164 posts were analyzed from five distinct tourism brands in Spain (dataset is in Spanish). Authors trained a regression model for predicting the number of likes and the number of comments a post will generate. For this, authors proposed as features the post richness (how many videos, pictures, links are included in the post), time frame (weekday and time of the publication), plus a couple of features associated to the size of the post (in characters) and the number of followers of the brand's page. Similar to the above-described research,  a few studies analyze the importance of the so-called contextual features (URLs, mentions, hashtags) to infer the number of replies a tweet may provoke \cite{Jenders:2013:APV:2487788.2488017, 5590452}. Finally, in the work described in \cite{yano2010s}, authors  model the relationship between the text of a political blog post and the number of comments that the post will receive. Authors approached the problem both as a regression problem and as a classification task. An interesting aspect of this work is that as features, authors employed a topic based representation (LDA) instead of metadata-based features. Given the nature of their data, they hypothesize that the nature of the topic contained in the post will influence the number of generated comments.

A common characteristic in previous research is the exclusion of text-based features (except for \cite{yano2010s}). Thus, contrary to previous research, our proposed framework incorporates three feature categories: stylistic, content-based, and behavioral. Our main hypothesis  establishes that the content of a post (what it says), as well as the style in how is written (how it say it), in combination with how the post is designed for interacting with the community (behavioral aspects) are important elements for accurately predicting the impact of a post.
We validate our proposal on a dataset with near 14,000 posts from ten different brands on Facebook Mexico, and compare our results against traditional metadata-based features. 

\section{Dataset}
\label{sect:corpus}

Given the lack of a standard corpus for evaluating impact prediction systems, we took on the task of collecting and standardizing a large dataset\footnote{The dataset is available in: \url{https://github.com/lyr-uam/CorpusReaccion}} of Facebook posts from different brands that have an important presence in Mexico\footnote{Compilation of the data was done from November 2018 to January 2019.}. Collected corpus represents a valuable resource, in a non-English language, that can be used for training and evaluating automatic systems that aim at predicting several customer's engagement metrics, specifically Facebook's reactions (i.e., Like, Love, Haha, Wow, Sad and Angry),  sharing amount, and the number of comments generated by a post. Table \ref{tab:DataSummary} summarizes the composition of the dataset. 

Under the columns \textbf{Num. of Posts}, we report the original (OG) number of collected posts and the resultant number of posts after filtering (FL) the data and eliminating those posts that were identified as useless. Particularly, we removed all the posts that fulfill any of the following conditions: \textit{i)} does not contain any text, \textit{ii)} does not have any reaction, and \textit{iii)} the only reaction contained is 'like'. At the end, a total of 871 posts were removed after applying the previous conditions. The first columns report some statistics regarding the number of \textbf{Reactions} (R), \textbf{Comments} (C), \textbf{Shares} (S), for each brand. Values below columns $|R|$, $|C|$, and $|S|$, represent the total number of reactions, comments, and shares, respectively. Values under the columns $\bar{x}$ and $\sigma$ indicates the average and standard deviation of reactions, comments, and shares for each post. It is worth mentioning that these statistics were computed with the FL version of the dataset. It is worth mentioning that these statistics were computed with the FL version of the dataset. In addition, keep in mind that these numbers may vary if the corpus is re-downloaded; since the date of compilation, posts could have generated more manifestations.


\begin{table*}[t]
    \centering
    \caption{Table shows the absolute number of reactions, comments and shares in the data set. Additionally, average and standard deviation values of these characteristics are shown} 
    
    \scriptsize
    \begin{tabular}{p{2cm}c@{~~}c@{~~}c@{~~~~~}c@{~~}c@{~~}c@{~~~~~}c@{~~}c@{~~}c@{~~~~~}cc}
    \toprule
     \multirow{2}{*}{\textbf{Brand's Name}}& \multicolumn{3}{c}{ \textbf{Reactions (R)}} &  \multicolumn{3}{c}{\textbf{Comments (C)}} &  \multicolumn{3}{c}{\textbf{Shares (S)}} & \multicolumn{2}{c}{\textbf{Num. of Posts}} \\
     \cmidrule(lr){2-4} \cmidrule(lr){5-7} \cmidrule(lr){8-10}\cmidrule(lr){11-12}
     & $|R|$ & $\Bar{x}$ & $\sigma$ & $|C|$ & $\Bar{x}$ & $\sigma$ & $|S|$ & $\Bar{x} $ & $\sigma$ & (OG) & (FL) \\
    \midrule
    Clash Royale ES  & 3,464,687 & 6209.12 & 11457.63 & 561,540 & 1006.34 & 2341.34 & 258,904 & 463.99 & 1495.29 & 561  & 558\\
    Canon Mexicana   & 2,316,406 & 2079.36 & 6626.56  & 112,280 & 100.79  & 265.89  & 426,502 & 382.88 & 1030.55 & 1157 & 1114\\
    Muy Interesante México & 14,900,267 & 6872.82 & 11314.93 & 258,214 & 119.10 & 296.41 & 4,801,967 & 2214.93 & 8436.72 & 2175 & 2168 \\
    Cinépolis & 28,074,276 & 14404.45 & 28790.81 & 2,784,917 & 1428.90 & 4179.64 & 8,165,961 & 4189.82 & 18652.82 & 1985 & 1949\\
    Discovery Channel &2,422,143 & 1446.06 & 2164.16 & 44,258 & 26.42 & 63.43 & 392,068 & 234.07 & 474.24 &1712 & 1675\\
    National Geographic &2,700,761 & 1548.60 & 3680.85 & 107,884 & 61.86 & 247.85 & 1,034,515 & 593.19 & 4643.73 &2076 & 1744\\
    Fisher-Price  & 3,550,516 & 4216.76 & 6053.21 & 155,811 & 185.05 & 400.89 & 249,981 & 296.89 & 709.63 &848 & 842\\
    Xbox México   & 4,697,179 & 2839.89 & 6360.91 & 479,033 & 289.62 & 749.36 & 513,323 & 310.35 & 860.45 & 1737 & 1654\\
    Nikon   & 829,560 & 673.34 & 1306.00 &  36,308 & 29.47 & 81.29 & 145,370 & 118.00 & 262.84 & 1357 & 1232\\
    Lacoste & 1,057,118 & 1478.49& 2559.50 & 10,005 & 13.99 & 36.79 & 37,037 & 51.80 & 266.68 & 914 & 715\\
    \midrule
    \textit{Total:}   & \textit{64,012,913} & \textit{-} & \textit{-} & \textit{4,550,250} & \textit{-} & \textit{-} & \textit{16,025,628} & \textit{-} & \textit{-} & \textit{14522} & \textit{13651}\\     
    \bottomrule
    \end{tabular}
    \label{tab:DataSummary}
\end{table*}

Observe in Table \ref{tab:DataSummary} that the brand with the highest number of reactions, comments, and shares is \textit{Cinépolis}. This brand is a very well known firm in Mexico, devoted to the movie theater business. The second place in the number of reactions and shares is held by  \textit{Muy Interesante México}. This is a firm mainly dedicated to science and technology diffusion. It is interesting to notice that even though \textit{Cinépolis} provokes a high number of manifestations from users, is not the brand that produces the most number of posts,  which is the case of \textit{Muy Interesante México} with the biggest number of posts. 

In Table \ref{tab:corpusStatistics} we show some basic statistics regarding the size of the corpus. The first three columns indicate the size of the collected data for each brand in terms of the number of tokens, the size of the vocabulary, and the lexical richness of the posts. Next two columns show the average number of tokens, and characters contained in every post of every brand. For the latter two, the standard deviation of these metrics is shown between parenthesis. 

From Table \ref{tab:corpusStatistics} we can remark that the brands with the largest number of tokens are \textit{National Geographic} and \textit{Discovery Channel}, both dedicated to promoting a great variety of programs related to ecology, wildlife, science, among others. Having a great number of tokens indicates that, in general, published posts from these brands are larger in terms of words per post. This phenomenon can be observed in the fifth column of Table \ref{tab:corpusStatistics} where it is possible to see the average number of tokens in the published posts. Lexical richness (LR) is a value that indicates how the terms from the vocabulary are used within a text. Is defined as the ratio between the vocabulary size and the number of tokens from a text ($LR=|V|/|T|$). Thus, a value close to 1 indicates a higher LR, which means vocabulary terms are used only once, while values near to 0 represent a higher number of tokens used more frequently (i.e., more repetitive). From our dataset, observe that the brands with the lowest LR values are \textit{National Geographic} and \textit{Cinépolis}, which means their produced posts employ a similar language. We hypothesize that this could be a marketing strategy since, for the case of \textit{Cinépolis}, allows them to reach a high number or manifestation in their posts in spite of being reiterative.

\begin{table*}[t]
    \centering
    \caption{This table shows the total number of tokens, vocabulary, and lexical richness of each brand's posts. Additionally, we show the average number of tokens, and characters for each post; between parenthesis the standard deviation is indicated}
     \scriptsize
    \begin{tabular}{lccccc}
    \toprule
    \multirow{2}{*}{\textbf{Brand's Name}} & \multicolumn{3}{c}{\textbf{Total number of:}} & \multicolumn{2}{c}{\textbf{Average number per post: }}\\
    \cmidrule(lr){2-4}\cmidrule(lr){5-6}
    & \textit{tokens}& \textit{vocabulary} & \textit{lexical richness} &\textit{tokens} ($\sigma$) & \textit{characters} ($\sigma$)\\
    \midrule
    Clash Royale ES (CR) & 14,531& 4,046& 0.27 &26.04 ($\pm$27.53) & 163.21 ($\pm$165.07)\\
    Canon Mexicana (CM)  & 21,885 & 5,006 & 0.22 &19.65 ($\pm$12.63) & 128.02 ($\pm$81.06) \\
    Muy Interesante México (MI)& 42,321 & 8,916 & 0.21 & 19.52 ($\pm$14.74) & 117.70 ($\pm$88.79)\\
    Cinépolis (CI)& 44,071 & 7,536 & 0.17 & 22.61 ($\pm$10.78) & 133.95 ($\pm$64.36)\\
    Discovery Channel (DC) & 44,659 & 10,862 & 0.24 & 26.66 ($\pm$12.94) & 158.09 ($\pm$75.28)\\
    National Geographic (NG) &46,039 & 6,988 & 0.15 & 26.40 ($\pm$13.33) & 153.16 ($\pm$73.32)\\
    Fisher-Price (FP) & 19,863 & 4,788 & 0.24 & 23.59 ($\pm$78.70) & 149.89 ($\pm$517.25)\\
    Xbox México (XM) & 27,639 & 5,283 & 0.19 & 16.71 ($\pm$7.21)  & 112.27 ($\pm$52.98)\\
    Nikon   (NK)   & 28,251 & 6,044 & 0.21 & 22.93 ($\pm$42.99) & 147.28 ($\pm$278.2)\\
    Lacoste  (LC)   & 13,455 & 3,570 & 0.26 & 18.82 ($\pm$24.75) & 128.06 ($\pm$155.12)\\
    \bottomrule
    \end{tabular}
    
    \label{tab:corpusStatistics}
\end{table*}

\subsection{Labeling methodology}
As we mentioned, our goal was to collect a data set for evaluating the performance of automatic methods for determining the impact of publishing a post on Facebook, in other words, anticipate the consumers' engagement. For this purpose, traditional engagement metrics were considered: reactions, comments, and sharing. 

Therefore, and inspired on the work of \cite{yano2009predicting, yano2010s}, we define the task of predicting consumer's engagement as the process of classifying whether a post will have higher (or lower) impact volume than the average seen in training data. Even though more fine-grained predictions are possible as well (e.g., predicting the absolute number of distinct reactions, the number of provoked comments, and the number of times is shared), our goal in this paper was not oriented to propose a methodology based on regression algorithms. Consequently, we define six binary classification problems, namely:  \textit{i)} comments ($|C|$), \textit{ii)} sharing ($|S|$), \textit{ii)} total reactions ($|R|$), \textit{iv)} positive reactions ($|R+|$), \textit{v)} negative reactions ($|R-|$) and, \textit{vi)} neutral reactions ($|R\odot|$). Each classification problem has categories \textit{high-impact} and \textit{low-impact}. 

The followed methodology for assigning each post's category, i.e., either  \textit{high-} or \textit{low-} impact, consists in the following steps: for each classification problem (i.e., the considered metrics), we compute the average value of metric $k$ among all the posts from the ten brands, this is referred as $\bar{x}_k$. Once we know the value $\bar{x}_k$, for each post contained in brand $i$, we review the value of metric $k$ in post $p_{i}$, thus, if $p_{i,k} > \bar{x}_k$ the category of the post is assigned to \textit{high-impact}, or \textit{low-impact} otherwise. This process represents a very straightforward approach for the problems of \textit{total reactions, comments}, and \textit{sharing}. However, for labeling \textit{positive}, \textit{negative} and \textit{neutral} reactions we acted as follows; we grouped as positive reactions the Like and Love responses, as negative reactions the Sad and Angry responses, and as neutral reactions the Wow and Haha responses.    
Table \ref{tab:corpusBalance} shows the number of instances on each category after the labeling process. As expected, the dataset has a much greater rate of low-impact volume posts for all the classification problems, potentially making the prediction task much harder problem.

\begin{table}[t]
    \centering
    \caption{Number of \textit{high-} and \textit{low-} impact instances for each problem}
    \begin{scriptsize}
    \begin{tabular}{l@{~~}c@{~}c@{~~}c@{~}c@{~~}c@{~}c@{~~}c@{~}c@{~~}c@{~}c@{~~}c@{~}c@{~}}
    \toprule
    \multirow{2}{*}{}& \multicolumn{2}{c@{~}}{\textbf{$|R|$}} & \multicolumn{2}{c@{~}}{\textbf{$|R+|$}}& \multicolumn{2}{c@{~}}{\textbf{$|R-|$}} & \multicolumn{2}{c@{~}}{\textbf{$|R\odot|$}}& \multicolumn{2}{c@{~}}{\textbf{$|C|$}}& \multicolumn{2}{c@{~}}{\textbf{$|S|$}}\\
    \cmidrule(lr){2-3}\cmidrule(lr){4-5}\cmidrule(lr){6-7}\cmidrule(lr){8-9}\cmidrule(lr){10-11}\cmidrule(lr){12-13}
    & \textit{high} & \textit{low}& \textit{high} & \textit{low}& \textit{high} & \textit{low}& \textit{high} & \textit{low}& \textit{high} & \textit{low}& \textit{high} & \textit{low}\\
    \midrule
    CR & 189 & 369  & 165 & 393  & 209 & 349  & 217 & 341  & 264 & 294  & 39  & 519\\
    CM & 100 & 1014 & 94  & 1020 & 43  & 1071 & 90  & 1024 & 78  & 1036 & 96  & 1018\\
    MI & 775 & 1393 & 790 & 1378 & 136 & 2032 & 374 & 1794 & 153 & 2015 & 824 & 1344 \\
    CI & 991 & 958  & 966 & 983  & 353 & 1596 & 729 & 1190 & 883 & 1067 & 745 & 1204\\
    DC & 109 & 1566 & 112 & 1563 & 110 & 1565 & 85  & 1590 & 9   & 1666 & 60  & 1615\\
    NG & 124 & 1620 & 132 & 1612 & 110 & 1634 & 53  & 1691 &  50 & 1694 & 115 & 1629\\
    FP & 248 & 594  & 266 & 576  & 15  & 827  & 31  & 811  & 119 & 723  & 43  & 799\\
    XM & 230 & 1424 & 230 & 1424 & 168 & 1486 & 155 & 1499 & 299 & 1355 & 92  & 1562\\
    NK & 24  & 1208 & 33  & 1199 & 16  & 1216 & 6   & 1226 &  12 & 1220 & 10  & 1222\\
    LC & 46  & 669  & 57  & 658  & 0   & 715  & 2   & 713  &  2  & 713  &  4  & 711\\
    \bottomrule
    \end{tabular}
    \end{scriptsize}
    \label{tab:corpusBalance}
\end{table}

\section{Proposed framework}
\label{sec:proposedframework}

\begin{figure*}[t]
    \centering
    \includegraphics[width=1.0\textwidth]{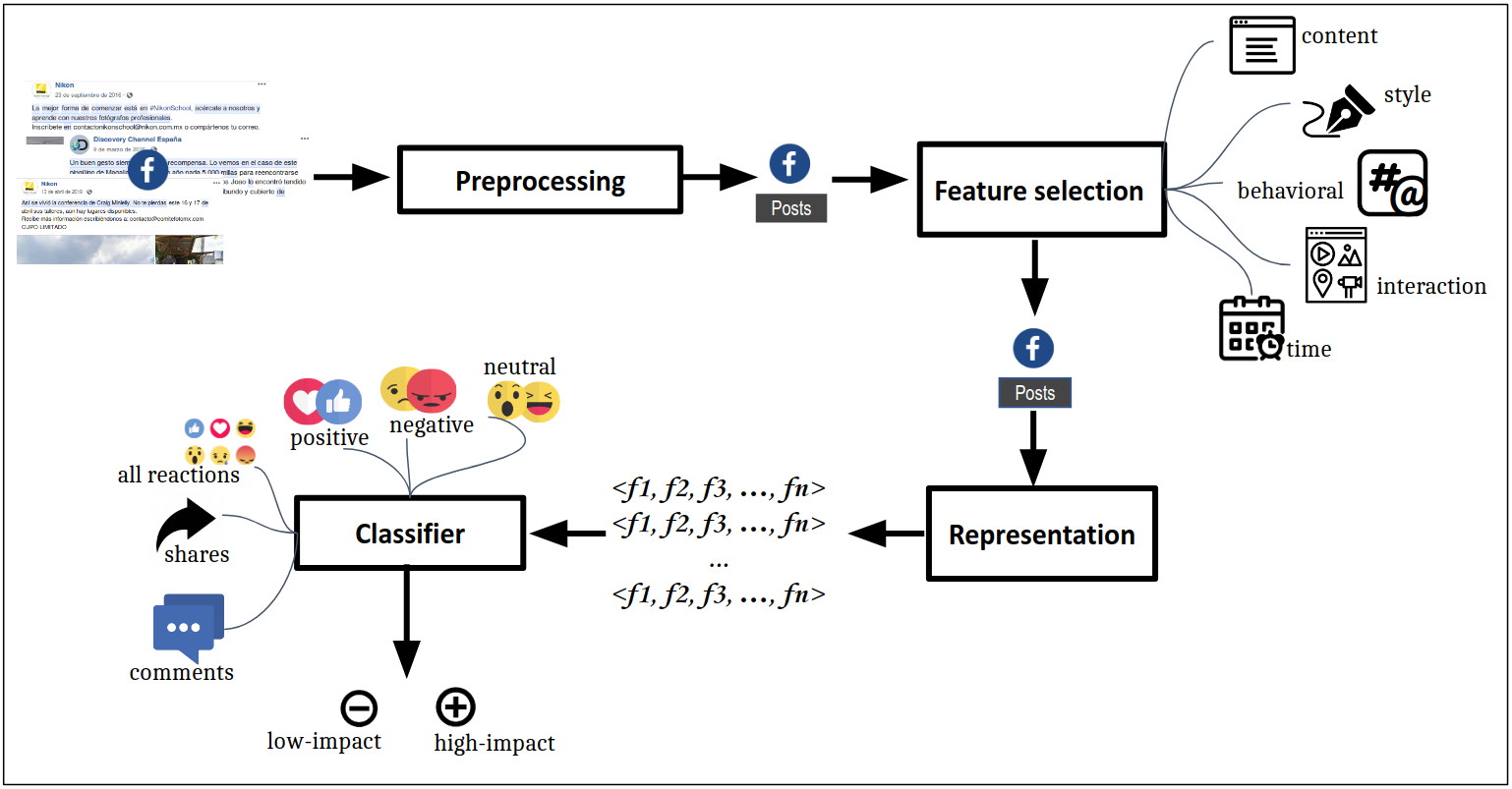}
    \caption{General framework of our proposed method.}
    \label{fig:general_framework}
\end{figure*}

Our general framework relies on the traditional pipeline of an automatic classification system. The classification problem is a learning problem, where the function $F(x) = y$ needs to be learned given a series of pairs $<x, y>$ where $x$ is an example of an instance and $y$ is the class of such example. Usually, $y \in Y$ and $|Y|$ is the total number of predefined class for a given classification problem.

Particularly, our goal is to learn six functions, one for each metric that are relevant to know the overall impact of a Facebook's post. Thus, our six classification problems are: impact of total number of comments, impact of number of shares, impact of total reactions, as well as, impact of positive reactions, impact of negative reactions, and impact of neutral reactions. And the predefined classes are high-impact and low-impact. 


The methodology used in all classification problems is the same as Figure \ref{fig:general_framework} presents. Each process in this figure is explained below:

\subsection{Preprocessing}
First, for each post ($x$) a preprocessing is performed. The general idea to this step is to estandarice the posts content to avoid having textual attributes with the similar semantic information. For instance, we do not care about all different URLs included in the posts, we only need to know that a post has an URL.

In this regard we replace all different  url, hashtag, emojis and mentions to unique tags such as \textit{<url>},\textit{ <hashtag>}, \textit{<emoji>}, and \textit{<mentions>}, respectively. 

\subsection{Feature selection}
The next process in Figure \ref{fig:general_framework} is a feature selection process. As we established, we want to include information about the \textit{what} is been posted as well as \textit{how} these posts are written. Consequently, in this process of our methodology we extracted the following types of features: Content-based that capture \textit{the what}, Style-based and Behavioral to capture \textit{the how}, and we include two more matadata-based features (as these are usually included in the previous works): Interaction and Time.

For the \textbf{content type}, we only considered single words as attributes. This feature can give us a general idea of \textit{what} is being saying in a post. The number of features of this type can vary accordingly to the vocabulary of the dataset. The \textbf{style type} features account for differences in the writing style of an author of that post. Ideally, a brand's post needs to have a similar writing style that aligns with such brand. In this regard, we included five features: the post's length (measure in words), the total number of upper-cased and lower-cased used, the total number of numerals and the total number of symbols (including punctuation marks and other non-alphanumeric symbols). To take into account the design for engagements, we selected the \textbf{behavioral type}. For this type we include four features: total number of emojis, total number of hashtags, total numbers of mentions and total number of links. 


Additionally, we include two types of metadata attributes: type of links included in the posts (we called this type \textbf{interaction}) and the time in which the post is written (\textbf{time}). Particularly, for the Interaction type we included five features: number of links to images, number of links to albums, number of links to videos and number of other links. For the Time feature we include the percentages of posts written at the same time if a given post, these percentages are compute independently for: hour, day, month and year.

\subsection{Representation}
Once we had selected the corresponding feature type, we represent each post in a multidimensional vector, where the number of dimensions correspond to the total number of features of a given representation.

The vectors are normalized to values between 0 and 1 to reduce the impact of differences between ranges of different type of features.

\subsection{Classification model}
The four phase of the general framework (see Figure \ref{fig:general_framework}) is to apply a learning algorithm for each classification problem. For this stage, we apply four of the widely algorithms used for text classification. At the same time, we selected one algorithm of 4 different families: Probabilistic (Naïve Bayes), Decisions Trees (DT), with kernel functions (SVM), and Instance-based (k-NN).

As we have mentioned, to provide an overview of the general impact of a post in the consumer, we generate six different prediction algorithms. At the end, the content manager of a given brand can determined the average impact given the predicted impact of comments, shares, total reactions, as well as, positive reactions, negative reactions and neutral reactions.
In the next section we describe the experiments performed as well as the obtained results.

\section{Experiments and results}
\label{sec:experiments}
To test method, we used the Filtered Dataset (FL in Table \ref{tab:corpusStatistics}) with a total of 13651 Facebook's posts. To evaluate the classification performance we used the F-score metric, and for all experiments we employ a 10 fold cross validation technique to compute the performance. Note that we do not make any distinctions of a posts for brand's name, since we wanted to build general predictors more that one predictor per brand.

One question we asked was if the combination of the \textit{what} plus the \textit{how} in the process of post's representation can be better at predicting the impact of our six metrics than using only features that answered the \textit{how}. With this in mind we performed two sets of experiments. In the first set, we used single attributes as representation for each post, each of these attributes have been used in previous works one way or another. The second set of attributes we include the content type of feature to investigate if in doing so we can improve the performance of each predictor.

\begin{figure}[tph!]
    \centering
    \includegraphics[width=0.6\textwidth]{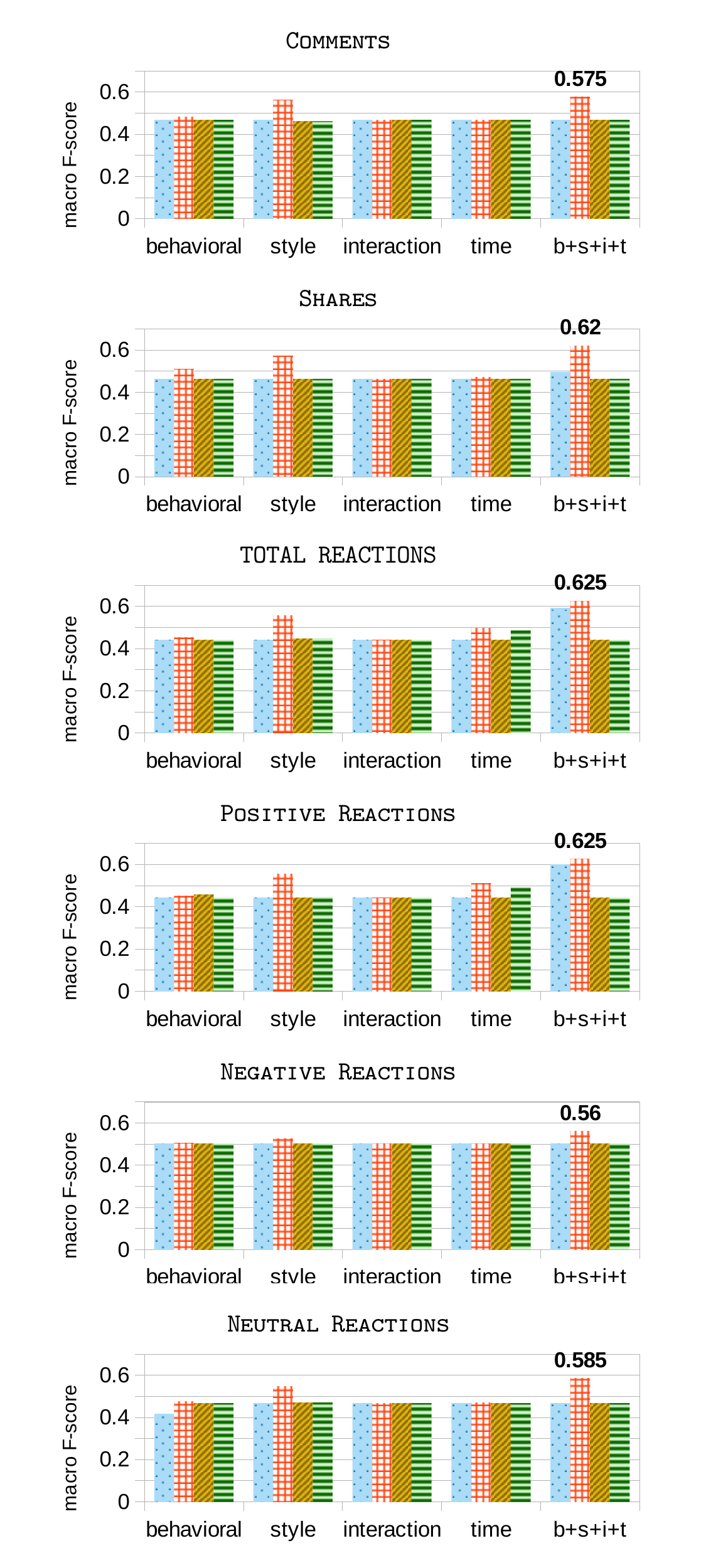} 
    \caption{Performance of our method for predicting the impact of brand's posts using our proposed features types independently. \textit{b+s+i+t} stands for the combination of behavioral, style, interactions and time features in the vector representation.}
    \label{fig:results_single}
\end{figure}

Figure \ref{fig:results_single} shows obtained results for the first set of experiments. It is important to mention that the size of the representation vector for each of these experiments is very small (between 4 and 5 features). One detail to notice in the Figure \ref{fig:results_single} is that the best classification algorithm for all problems is Decision Trees, which makes sense given the small number of attributes used as post's representation. Also, we can observer that the style attribute alone, is the second best predictor of each problem. However, the best performance outcome happens when a combination of the four type of attributes is used (\textit{b+s+i+t}). Among the less useful set of features are the metadata-based ones: interactions and time, where all instances were classify as the majority class (i.e., low-impact); wherefore, these two sets of features are not used in the second set of experiments. 

So far, Figure \ref{fig:results_single} show very consistent results for all problems; but nevertheless, minor aspects are worth mentioning. For instance, the most difficult classification problem is predicting the impact of negative reactions. However, one of the better performance is in predicting the impact of positive reactions. These differences may be due to the slightly less unbalance dataset for positive reactions in contrast with the dataset for negative reactions.

\begin{figure}[tph!]
    \centering
    \includegraphics[width=0.6\textwidth]{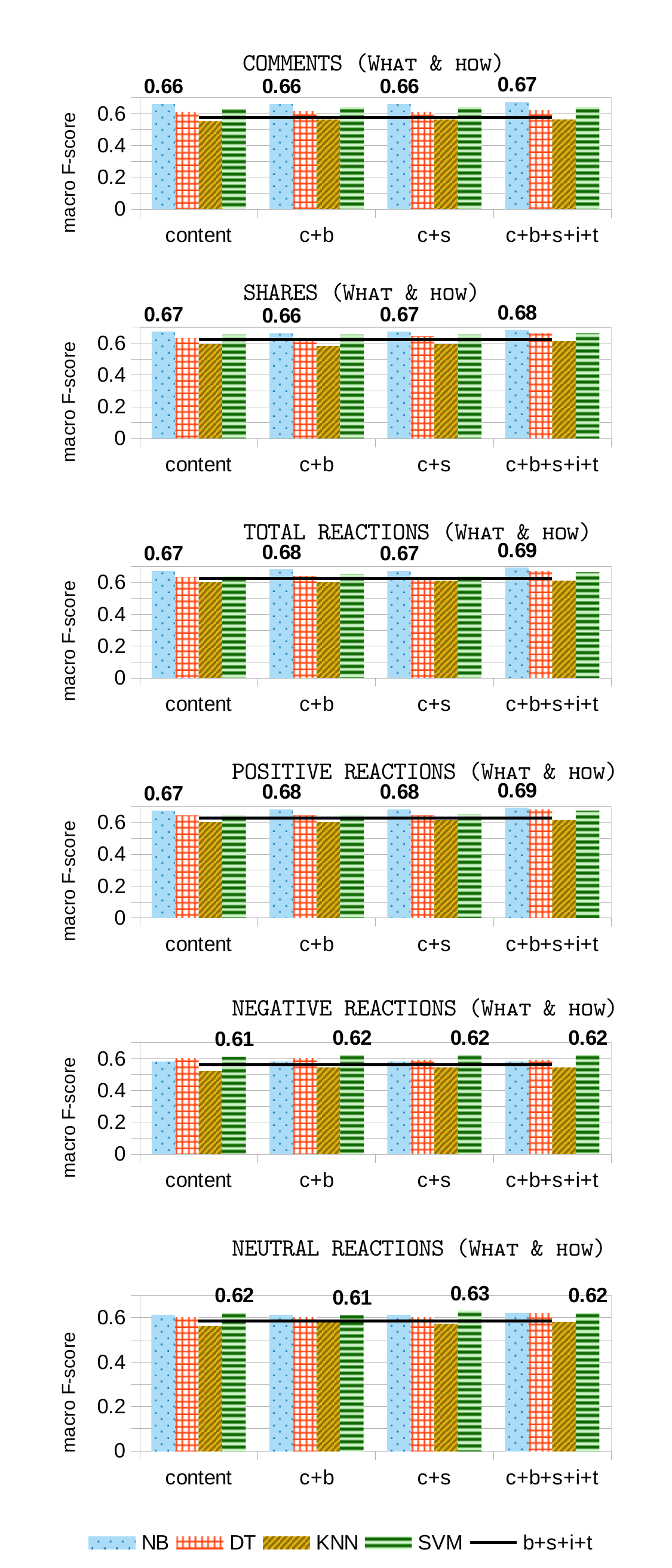} 
    \caption{Performance of our method for predicting impact of brand's posts using our proposed features types in combination with the text (content feature). \textit{c}, \textit{b}, \textit{s}, \textit{i} and \textit{t} stands for content feature, behavioral feature, style feature, interaction and time features, respectively.}
    \label{fig:results_multiple}
\end{figure}

In Figure \ref{fig:results_multiple} results of the second set of experiments can be seen. For including the text, we used a bag-of-word approach to represent each post. We used only the 10000 tokens more frequent in each problem. The black solid line in each graph indicates the best performance of the previous set of experiments (i.e., from Figure \ref{fig:results_single}). As a general view, we notice that for all problems using the content feature (alone or in combination with other type of feature) outperformed the best results using only single attributes. Another aspect to note is that, on one hand, the best learning algorithm for four out of six problems is the probabilistic one. Support Vector Machines, on the other hand, is also the best algorithm predicting the impact of negative and neutral reactions. Both algorithms had been successful in text classification in the past. 

As in results from Figure \ref{fig:results_single}, the poorest performance was in the prediction of impact of negative and neutral reactions. That is, those two problems are very difficult to solve. On the contrary, the best overall performances were obtained for predicting the impact of total reactions and positive reactions, follow by predicting the impact of shares and comments.

In all problems, the best performance was obtained using the combination of all our proposed features: the what and the how plus the metadata information. On one hand, the small different in the performance of using only the content feature (the what) with the best results, particularly for predicting the impact of Comments and Shares, gives us some clue of the importance of the content in predicting our six variables. On the other hand, for predicting reactions (total or positive reactions, particularly) the improvement in the performance when combining content with behavioral features, tends to improve the performance. This, can means that using the number of occurrences of hashtag, emojis, mentions and links is important to predicting the impact of a post. This type of feature was included to give some information of the social media lingo used when communicating some information. Seems like including this kind of tokens helps to reach the consumers and induce them to express their feeling towards the brand.






\subsection{Qualitative results}
\label{subsec:qualitativeResults}
Aside from the prediction tasks such as above, the proposed approach itself can be informative for people in charge of designing marketing strategies. As stated so far, our proposed framework is able to determine the impact of publishing a post on Facebook. Given that part of our goals was to design a generic impact prediction method, i.e., not brand dependent, our approach allows us to envisage characteristics from high and low impact posts. 

\begin{figure*}

  \includegraphics[width=1.0\textwidth]{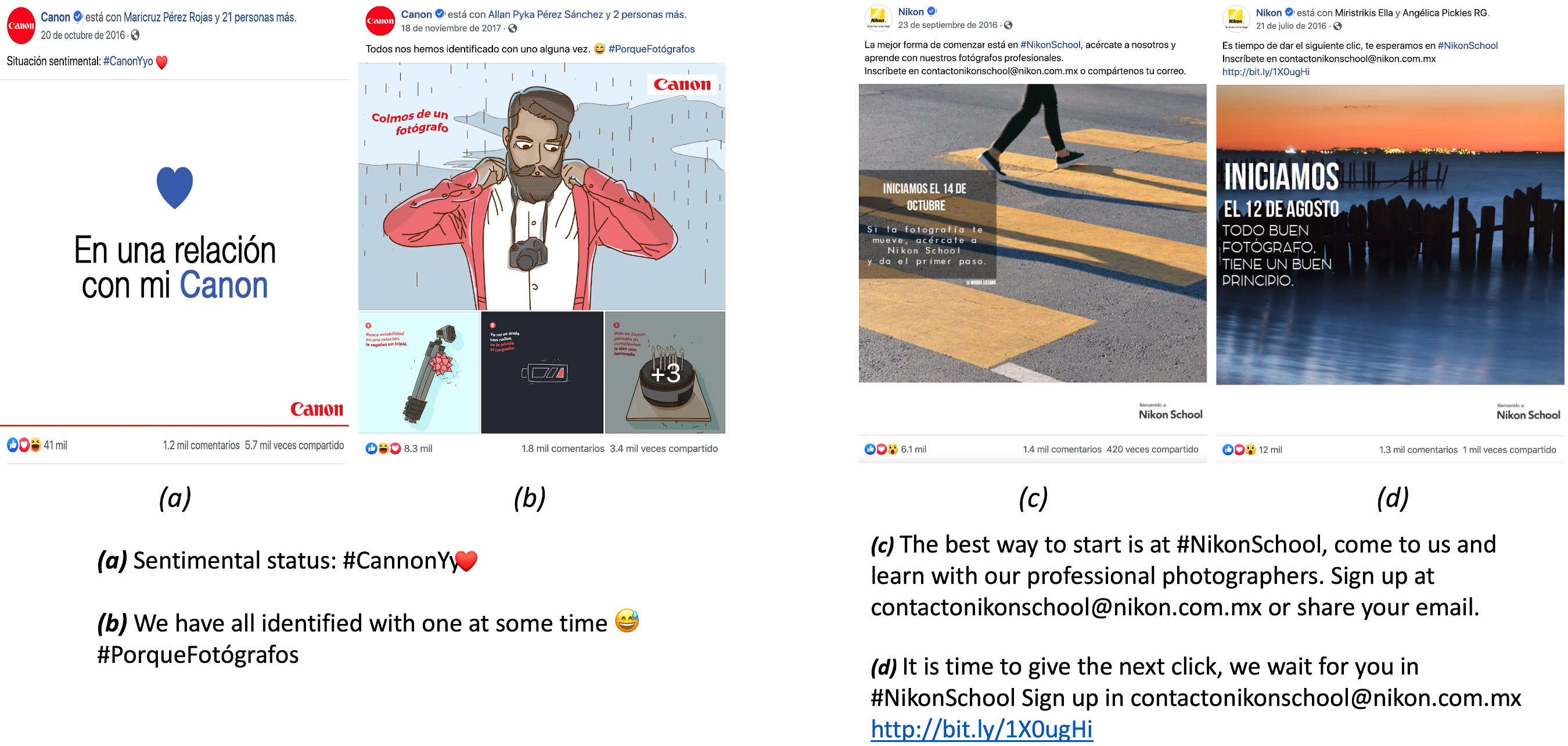}\\

  \includegraphics[width=1.0\textwidth]{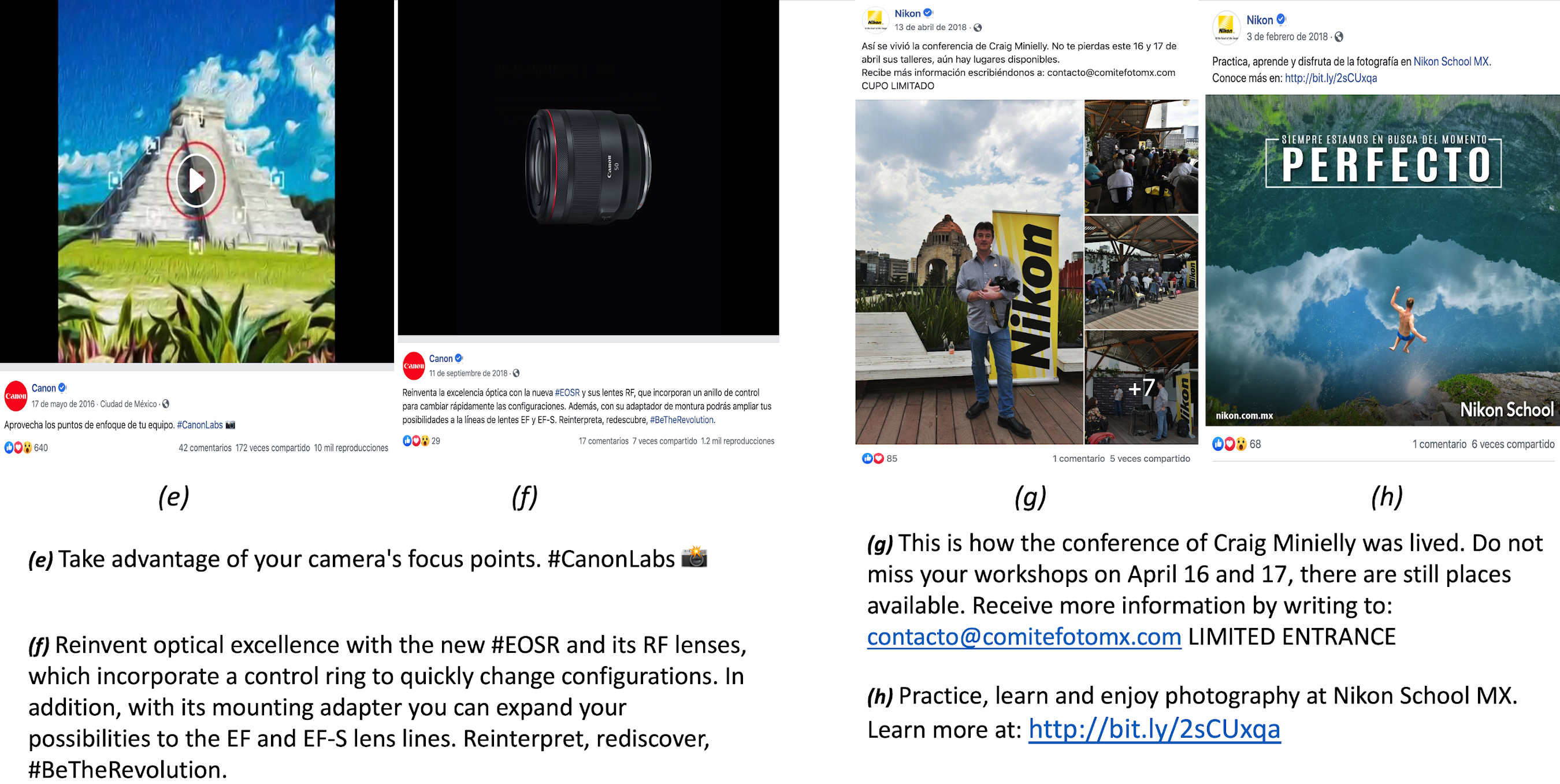}

\caption{Examples of high-impact (\textit{a, b, c}, and \textit{d}) and low-impact (\textit{e, f, g}, and \textit{h}) published posts extracted from the \textit{Cannon Mexicana} and \textit{Nikon} }
\label{fig:examplesHighLow}
\end{figure*}

In order to exemplify the type of information that can be obtained with our proposed method, we retrieve eight examples (four high-impact, and four low-impact posts), and analyze its characteristics. In Figure \ref{fig:examplesHighLow} we show two high-impact posts (\textit{a} and \textit{b}), and two low-impact posts (\textit{e} and \textit{f}) from \textit{Cannon Mexicana}. Similarly, we retrieved two high-impact posts (\textit{c} and \textit{d}) and two low-impact posts (\textit{g} and \textit{h}) from \textit{Nikon's} Facebook page.  

Given the nature of these two brands, we found interesting to analyze its publications. As it is known, these two firms compete in the field of photography, they both promote photography courses, professional photography equipment, etc. If we observe Table \ref{tab:DataSummary}, notice that Nikon publishes a bit more posts than Cannon Mexicana (1,357 \textit{vs.} 1,157). However, Cannon Mexicana has a significantly greater number of reactions, comments, and shares than Nikon; for example, Cannon has more than 2 million reactions while Nikon has barely 829,560.  After examining their most representative posts (Figure \ref{fig:examplesHighLow}), we notice the following: \textit{i)} High-impact Cannon's posts have a more juvenile way for interacting with their customers, they employ emojis, drawings, as well as less-formal language; \textit{ii)} contrastingly, Nikon uses a more formal style of writing, and their posts refer (mainly) to photography courses, while for Cannon, their posts refer to photographers activities or situations.  

With respect to the low-impact posts for both firms, it is interesting that for Cannon, their less popular posts talk about technicalities of the cameras, such as focus points and lens' characteristics. A similar phenomenon occurs for the case of Nikon, where their less popular posts talk about the results of a workshop. Thus, as a preliminary result from this analysis, we could conclude that Nikon needs to produce less formal posts in order to reach a higher level of consumer engagement activities, in other words, change the way they use emojis, hashtags, mentions or links in their published posts.  

We performed a similar analysis between the high- and low-impact posts from \textit{Discovery Channel} and \textit{National Geographic}. Due to the lack of space, we don't show the obtained most relevant posts. However, for this particular case, we found that for both brands, their less popular publications always refer to TV programming of their respective channels. Regarding their most popular posts, we found that every time these brands publish something related to science and technology diffusion, customers engage positively.


\section{Conclusions and future work}
\label{sec:conclusions}
This paper focused on proposing a novel framework for anticipating the impact of publishing a post on a company's Facebook page. Our main hypothesis establishes that if an automatic classification algorithm is able to accurately model the what and the how a post should be written, then it will be possible to predict its impact, i.e., its consumer engagement level. Thus, our proposed approach incorporates features that are able to capture content, style, and behavioral characteristics from posts. 

In order to validate our hypothesis, and given the lack of a standard corpus for evaluating this type of approaches, we took in the task of collecting and standardizing a large dataset of Facebook posts from different brands in Mexico. The collected corpus represents a major contribution of this work, and aims at providing resources for future research work in non-English languages. Accordingly, we evaluated our proposed approach in predicting traditional engagement metrics, such as reactions (total reactions, positive, negative, and neutral), comments, and sharing. We performed experiments on our collected dataset, which contains more than 13,000 posts from ten different brands on Facebook Mexico, and compare our results against traditional metadata-based features. Obtained results indicate that \textit{what} and \textit{how} the companies write, in combination with some traditional metadata-based features, allows to obtain the best performance. A qualitative analysis allowed us to observe what are the aspects our proposed model is learning. For instance, we could notice that for some particular brands, competing in the same market, their behavior (i.e., the use of emojis, hashtags, or mentions), in combination with the topics of the post,  are very important for improving costumers engagement.

Some relevant advantages of the proposed method are: is a language-independent approach, is not biased towards a specific brand or product type, and allows to obtain relevant insights that could be beneficial for community managers providing them some interesting knowledge. 

Several ideas arise from this initial research for future work. First, the proposed model could be enriched with other stylistic and content features. For example, character n-grams are known for providing valuable stylistic information. Regarding content, we plan to incorporate some topic-based features, such a LDA or second order representations. Finally, there exist some evidence on the relevance of detecting the post's sentiment as a feature, we plan to evaluate how beneficial could be to incorporate this type of features in our framework. 

\section*{Acknowledgements}
Érika Rosas-Quezada was partially supported by the CONACyT Thematic Networks program (RedTTL Language Technologies Network) with project numbers: 281795 and 295022; she also wants to thank the Communication Sciences and Design Division from UAM Cuajimalpa, Mexico. Ramírez-de-la-Rosa wants to thank UAM Cuajimalpa for the support given to this research. And Villatoro-Tello was partially supported by ADOBE project, from Idiap Research Institute, Switzerland, and by the Information Technologies Department from UAM Cuajimalpa, Mexico. 


\bibliographystyle{abbrv}
\bibliography{referencias}

\end{document}